\documentstyle[aps,prb,multicol,epsf]{revtex}
\input psfig
\draft
\begin{document}
\date{July, 1st,1998}
\title{Magnetic Bound States in dimerized Quantum Spin Systems}
\author{P. Lemmens, M. Fischer, M. Grove, G. Els, G. G\"untherodt}
\address{2. Physikalisches Institut, RWTH-Aachen, 52056 Aachen, Germany}
\author{M. Weiden, C. Geibel, F. Steglich}
\address{MPI CPFS, 01187 Dresden, Germany}
\maketitle
\begin{abstract}

Magnetic bound states are a general phenomenon in low dimensional
antiferromagnets with gapped singlet states. Using Raman
scattering on three compounds as dedicated examples we show how
exchange topology, dimensionality, defects and thermal
fluctuations influence the properties and the spectral weight of
these states.

\end{abstract}

One-dimensional antiferromagnets with gapped singlet spin states
have attracted much attention in recent years from both
theoretical and experimental points of view. Such singlet ground
states are realized in the dimerized phase of spin-Peierls systems
(T$<$T$_{SP}$) or in alternating chain compounds. The magnetic
excitation spectrum of a dimerized spin chain contains a triplet
branch at $\Delta_{01}$, a corresponding two-particle continuum of
triplet excitations starting at 2$\Delta_{01}$, and well defined
magnetic bound states \cite{uhrig,bouz}. The latter consist of
strongly interacting triplet excitations. Above the high energy
cutoff 2$\Delta_{01}$ the bound states dissolve in the
two-particle continuum of propagating triplet excitations.

In 1D spin models (spin chains) with frustrated next-nearest
neighbor interaction $J_2= 0.24 J_1$, acting as a binding
potential, the existence of a singlet bound state at
$\sqrt{3}\Delta_{01}$ and of a further triplet bound state for
higher values of $J_2$ \cite{bouz} are predicted. Spin ladders
(half-filled) interpolate in some sense between one- and
two-dimensional quantum spin systems. Here similar magnetic bound
states should exist \cite{kotau}. Summarizing, a study of magnetic
bound states in quantum spin systems with a gapped excitation
spectrum gives valuable insight into the low energy fluctuations
which govern the physics of these systems.

{\bf I. CuGeO$_3$}\\ In the compound CuGeO$_3$ (T$_{SP}$=14K)
\cite{hase} a magnon-like triplet state,
$\Delta_{01}$=16.8~cm$^{-1}$, with a second, similar gap to the
continuum is observed in neutron scattering experiments for
T$<$T$_{SP}$ \cite{nishi}. The dispersion of this mode points to a
frustrated exchange along the CuO$_2$-chain direction and a still
considerable exchange perpendicular to this direction. In Raman
scattering experiments the singlet bound state is observed as a
sharp and asymmetric low energy mode at
$\Delta_{00}$=30~cm$^{-1}\approx$1.79$\Delta_{01}$ (see Fig. 1a)
\cite{lemmens}. It shows up only with incident and scattered light
polarizations parallel to the chain direction. These polarization
selection rules for a 1D spin system are consistent with the spin
conserving nature of the exchange light scattering mechanism
($\Delta$s=0). The intensity of the bound state increases very
slowly with decreasing temperature below T$_{SP}$ compared with
the intensity of the spin-Peierls induced phonon modes. Close to
T$_{SP}$ the frequency its is reduced and its linewidth enormously
broadend. In substitution experiments the intensity of the singlet
bound state drops drastically \cite{lemmens}. Instead of the
singlet bound state a new bound state ("dopand bound spinon") on
the defect site is induced at an energy comparable to
$\Delta_{01}$ \cite{els}.

{\bf II. $\alpha^\prime$-NaV$_{2}$O$_{5}$}\\ The compound
$\alpha^\prime$-NaV$_{2}$O$_{5}$ (T$_{SP}$=34K)
\cite{isobe,weiden}, now recognized as a quarter-filled spin
ladder \cite{smolinski}, is formed by layers of VO$_{5}$ pyramids
with mixed V$^{4+}$ (s=1/2) and V$^{5+}$(s=0) ions and a dominant
superexchange path via the oxygen plaquettes along the ladder
direction (b-axis). For T$>$T$_{SP}$ this system can be mapped on
an effective spin chain with the spins residing on V-O-V orbitals
\cite{smolinski}. For T$<$T$_{SP}$ a singlet-triplet gap
$\Delta_{01}$=60~cm$^{-1}$ was determined from magnetic
susceptibility measurements \cite{weiden}. The magnon-like triplet
branches investigated in neutron scattering experiments with a
steep dispersion along the b-axis (ladder direction) show a
bonding-antibonding like splitting with a maximum separation of
23.5~cm$^{-1}$ for k-vectors perpendicular to the ladder
\cite{fuji}. Three new modes (67, 107, 134~cm$^{-1}$) appear in
Raman scattering that have the same variation of intensity,
linewidth and frequency as function of temperature as the singlet
bound state in CuGeO$_3$ (see Fig. 1b) \cite{prl}. While the mode
at 107~cm$^{-1}$ is observed with light polarizations parallel to
the ladder direction, the other two modes are nearly unpolarized
within the ab-plane. The mode at 67~cm$^{-1}$ shows no splitting
in an applied magnetic field, thus excluding a one-magnon
scattering mechanism. The energy of this mode is extremely
renormalized compared to 2$\Delta_{01}$. Furthermore, the
separation between the bound states is comparable to the
bonding-antibonding splitting of the triplet branch. Therefore the
multiplicity of the bound states must be due to a dynamic
formation of the underlying spin dimers. A static dimer-dimer or
interchain interaction would only lead to one dispersing singlet
bound state. To allow for unpolarized light scattering diagonal
oriented dimers on the ladder should compete with parallel and
perpendicular dimer orientations. These degrees of freedom are
allowed since for T$<$T$_{SP}$ two distinguishable V-sites
(V$^{4+}$, V$^{5+}$) exist that may be disordered on the two legs
of the ladder \cite{prl}. The new ground state and the magnetic
bound states are then a result of a dynamical quasi-Jahn-Teller or
RVB model depending on whether spin-phonon coupling or electronic
correlations play the dominant role.

{\bf III. (VO)$_2$P$_2$O$_7$}\\ Neutron scat\-tering
in\-vesti\-gati\-ons on (VO)$_2$P$_2$O$_7$ show\-ed two strong\-ly
dis\-per\-sing trip\-let modes ($\Delta_{01}$=25~cm$^{-1}$ and
$\Delta_{11}$=46~cm$^{-1}$ at q=0) as the magnetic triplet bound
states of an alternating spin chain compound \cite{ger}. However,
recently these results were discussed within a 2D spin model with
an additional diagonal coupling \cite{newbouz}. In Raman
scattering experiments a shoulder of scattering intensity is
observed at 47~cm$^{-1}$, an energy corresponding to
2$\Delta_{01}$ (see Fig.1c) \cite{grove}. This intensity shows up
with light polarizations both parallel and perpendicular to the
dominant exchange path. It is not observed in crossed
polarization. This excludes two-magnon scattering of a usual 2D
spin system. Furthermore, the intensity of this shoulder as
function of temperature shows again a gradual increase with
decreasing temperature for T$<$75K$\approx$2$\Delta_{01}$ which
allows an assignment of this shoulder to a singlet bound state
with very small binding energy.

Using light scattering, the properties of magnetic bound states in
three different spin systems with a gapped singlet ground state
were investigated and compared with the corresponding neutron
scattering results. These magnetic bound states consist of triplet
excitations that are bound with respect to the "free" two-particle
continuum of states above 2$\Delta_{01}$. While their energy and
multiplicity are given by frustration and exchange topology of the
investigated spin system, their spectral weight is strongly
reduced by thermal fluctuations and nonmagnetic substitutions.

Financial support by DFG through SFB 341 and SFB 252, BMBF/Fkz
13N7329 and INTAS 96-410 is gratefully acknowledged. Samples of
(VO)$_2$P$_2$O$_7$ were kindly provided by B.C. Sales.

\begin{figure}
\centerline{\psfig{file=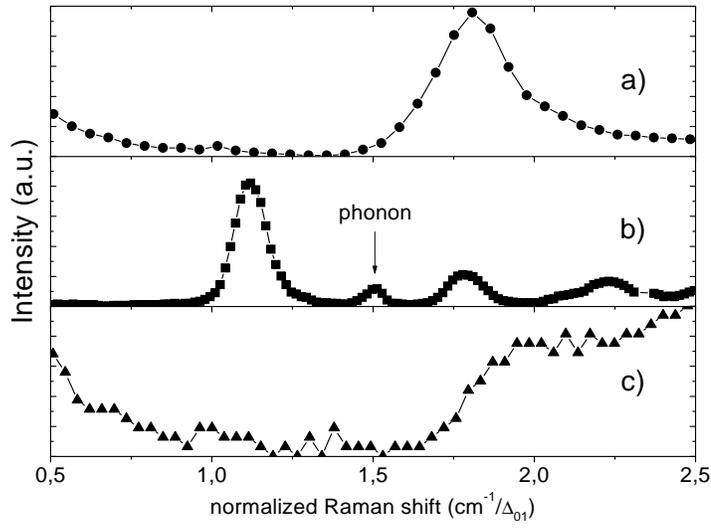,width=14.5cm}} \caption{Raman
spectra of a) CuGeO$_3$, b) $\alpha\prime$-NaV$_{2}$O$_{5}$ and c)
(VO)$_2$P$_2$O$_7$ at T=5~K on an energy scale (Raman shift)
normalized by $\Delta_{01}$. Light scattering polarizations are
parallel to the dominant exchange path.}
\end{figure}

\end{document}